\documentclass[12pt, a4paper]{article}
\usepackage[cp866]{inputenc}
\usepackage[dvips]{graphicx}
\usepackage{amsmath}
\usepackage{amssymb,latexsym}
\usepackage{amsfonts}
\begin{document}
\begin{center}
\bf{NEWTON'S SECOND LAW AND THE CONCEPT OF RELATIVISTIC MASS\\[5mm]}
\end{center}
\begin{center}
V. A. Pletyukhov \\
Brest State University, Brest, Belarus \\[3mm]
12.02.2018
\end{center}
\hspace{2.0cm}
 \begin{abstract}
In this work we discuss different interpretations of mass in the relativistic dynamics. A new way to introduce mass is proposed. Our way is based on the relativistic equation of motion expressed in the form of the Newton's second law. In this approach mass appears as a tensor, not as a scalar. The tensor mass allows us simply to describe anisotropic character of inert features of a relativistic object.
  \end{abstract}
\hspace{1.0cm}

After  creation of the Special  Relativity theory  more than 100
years have passed. On the fundamental of this theory among the
 professionals there exists a complete unity of opinions.
Nevertheless, on some particular issues there are still
differences in views, which from time to time give rise to heated
debates. As an example, one can refer to the publication of the
famous Russian theoretical physicist Academician L.B. Okun \cite{Okun}.

The author of this article has been teaching the theory of
relativity for more than 40 years to students and undergraduates
of the Faculty of Physics and Mathematics of Brest University. Our
point of view on the relativistic mass differs from the position
of both the esteemed academician and the opponents criticized by
him. I would like to share my thoughts with readers who are
interested in this issue.

Let us start with the most trivial things. As is known, the
classical equation of motion for a point body has the form
$$
{\frac{d}{dt}} (m\mbox{\boldmath $v$}) = \mbox{\boldmath $F$}
\qquad  \text{or} \qquad {\frac{d \mbox{\boldmath $p$}} {dt}} =
\mbox{\boldmath $F$}  \quad (\mbox{\boldmath $p$}=m\mbox{\boldmath
$v$}), \eqno(1)
$$
where  $m$ is an invariant quantity, it does not depend on   the
velocity  of the body and is  called  a mass. It is also assumed
that the mass has the property of additivity and  can be
considered as a measure of the amount of matter contained in the
body. The independence of the mass on  time allows us to rewrite
(1) as follows:
$$
\mbox{\boldmath $F$} = m {\frac{d \mbox{\boldmath $v$}}{dt}} =
m\mbox{\boldmath $a$}. \eqno(2)
$$
Equation (2) is called the Newton's second law. The mass in (2) is
just a   proportionality coefficient between force and
acceleration. In this role, it characterizes the inertia In
properties of the body, i.e. its ability to acquire a definite
acceleration under the action of a certain force.  In this sense,
they say that the mass is a measure of the inertness of the body.
Often it is called "inertial mass".

Thus, the mass in classical mechanics exhibits  three main
characteristics: invariance, additivity, and  measure of inertia.
They are  its "visiting card" \hspace{1mm} in the theory. Here
they peacefully coexist, complementing each other.

The situation is different in relativistic dynamics. The
three-dimensional relativistic equation of motion has the form
$$
{\frac{d}  {dt}} \left( {\frac{m\mbox{\boldmath
$v$}}{\sqrt{1-\beta^{2}}}} \right) = \mbox{\boldmath $F$}.
\eqno(3)
$$
Or, if we introduce the quantity
$$
M={\frac {m } {\sqrt{1-\beta^{2}}}} \eqno(4)
$$
and the relativistic momentum
$$
\mbox{\boldmath $p$} = {\frac{m\mbox{\boldmath $v$}}
{\sqrt{1-\beta^{2}}}} = M\mbox{\boldmath $v$}, \eqno(5)
$$
we can write (3) in a form analogous to (1):
$$
{\frac{d} {dt}} \left(  M\mbox{\boldmath $v$} \right) =
\mbox{\boldmath $F$}, \qquad {\frac{d \mbox{\boldmath $p$}}{dt}} =
\mbox{\boldmath $F$}. \eqno(6)
$$
The invariant quantity $m$ in equation (3) can no longer be
regarded as a measure of inertia. Proceeding from this, some of
physicists (see, for example, \cite{FeynmanL}) believe  that the role of mass
in relativistic dynamics is performed by the quantity (4), which
replaces  the classical mass in the relativistic equation of
motion. However, it is obvious that the quantity $M$ also is not a
measure of inertia, since in the general case
 $\textbf{F}\neq M\textbf{a}$.

The question arises whether it is correct to speak  about the
existence in the relativistic dynamics of a quantity that can
claim to be a measure of inertia in the above formulated sense. To
answer this question, we will to elucidate the  question --
whether equation (3) can be represented in a form analogous to
Newton's second law. To do this, let us consider the derivative
term in the left-hand side of equation (3):
$$
{\frac{d} {dt}} \left( {\frac{m \mbox{\boldmath $v$}}
{\sqrt{1-\beta^{2}}}} \right) = {\frac{m} {\sqrt{1-\beta^{2}}}}
{\frac{d\mbox{\boldmath $v$}} {dt}} + {\frac{m\mbox{\boldmath $v$}
\left(\mbox{\boldmath $v$}, {\frac{d\mbox{\boldmath $v$}} {dt}}
\right) } { c^{2} \left( 1-\beta^{2} \right)^{3/2}}}=
$$
$$
={\frac{m}{\sqrt{1-\beta^{2}}}} \left( 1+ {\frac {\mbox{\boldmath
$v$} \cdot \mbox{\boldmath $v$} } { c^{2} \left( {1-\beta^{2}}
\right)
     }
   }
\right)\mbox{\boldmath $a$}  = {\frac{m}{\sqrt{ 1-\beta^{2} } }}
\left( 1 + {\frac{\mbox{\boldmath $\beta$}\cdot \mbox{\boldmath
$\beta$}}   {1-\beta^{2}}}\right) \mbox{\boldmath $a$}. \eqno(7)
$$
Here  $ \mbox{\boldmath $\beta$} = {\frac{\mbox{\boldmath $v$}}
{c}}$, $ \left(\mbox{\boldmath $v$}, {\frac{d\mbox{\boldmath $v$}}
{dt}} \right) = \left( \mbox{\boldmath $v$},\mbox{\boldmath $a$}
\right)$ is a scalar product of vectors; $\mbox{\boldmath $v$} \cdot
\mbox{\boldmath $v$}$, $\mbox{\boldmath $\beta$}\cdot
\mbox{\boldmath $\beta$}$ is a diad product of vectors.
If we represent the vectors  \mbox{\boldmath $F$}  and \mbox{\boldmath $a$}   in the form of columns
$$
\mbox{\boldmath $F$} =
 \left(\begin{array}{ccc}
F_{1}\\
F_{2}\\
F_{3}\\
\end{array}\right), \qquad
\mbox{\boldmath $a$} =
 \left(\begin{array}{ccc}
a_{1}\\
a_{2}\\
a_{3}\\
\end{array}\right),
\eqno(8)
$$
then the diad  $\mbox{\boldmath $\beta$}\cdot \mbox{\boldmath $\beta$}$ represents a matrix
$$
\mbox{\boldmath $\beta$}\cdot \mbox{\boldmath $\beta$} =
\left(\begin{array}{ccc}
\beta_{1}^{2} &\beta_{1}\beta_{2} &\beta_{1}\beta_{3}  \\
\beta_{2}\beta_{1} & \beta_{2}^{2} &\beta_{2}\beta_{3}    \\
\beta_{3}\beta_{1} & \beta_{3}\beta_{2}    & \beta_{3}^{2}
\end{array}\right). \eqno(9)
$$
In the transformations (7), the dyad property is used:
 $(\mbox{\boldmath $a$}\cdot \mbox{\boldmath $b$})\mbox{\boldmath $c$} =
 \mbox{\boldmath $a$}(\mbox{\boldmath $b$}, \mbox{\boldmath $c$})$.

With (7) in mind, eq.  (3) reads
$$
\mbox{\boldmath $F$} = \mu \mbox{\boldmath $a$}, \eqno(10)
$$
where
 $\mbox{\boldmath $F$}$ and   $\mbox{\boldmath $a$}$ stand for 3-vector columns
of the type (8),  and $\mu$ is a  $3\times3$ matrix
$$
\mu = {\frac{m} {\sqrt{1-\beta^{2}}}} \left( 1 +
{\frac{\mbox{\boldmath $\beta$}\cdot \mbox{\boldmath $\beta$}}
{1-\beta^{2}}}\right). \eqno(11)
$$
In the index notation we will have
$$
F_{i}=\mu_{ij}a_{j}, \eqno(12)
$$
where
$$
\mu_{ij} =  {\frac{m} {\sqrt{1-\beta^{2}}}} \left( \delta_{ij} +
{\frac{\beta_{i} \beta_{j}}   {1-\beta^{2}}}\right) \eqno(13)
$$
and  we mean summation over repeated indices.

Equations (10), (12) can be reversed, namely:
$$
\mbox{\boldmath $a$} = \mu^{-1}\mbox{\boldmath $F$}, \qquad
a_{i}=\mu_{ij}^{-1} F_{j}, \eqno(14)
$$
where the inverse matrix $\mu^{-1}$ has the form
$$
\mu^{-1} = {\frac{\sqrt{1-\beta^{2}}}{m}} \left( 1 -
{\mbox{\boldmath $\beta$}\cdot \mbox{\boldmath $\beta$}}\right),
\qquad \mu_{ij}^{-1} = {\frac{\sqrt{1-\beta^{2}}}{m}} \left(
\delta_{ij} - \beta_{i}\beta_{j} \right).
\eqno(15)
$$

It is obvious that the matrix $\mu$, as a
 "proportionality coefficient" \hspace{1mm}
 between the vectors of force and acceleration
 reflects  fully and accurately  the meaning of the concept
 of "measure of inertia"  \hspace{1mm} in relativistic dynamics.
 The matrix nature of the mass as a measure of inertia
  means that the inertial properties of the relativistic body are not isotropic.
  This non-isotropy is due to the fact that
  the moving body automatically creates in space at each instant of time
  a selected direction which coincides with the direction of the velocity of the body.
  Therefore, we can say that non-isotropy has a kinematic character
   and is manifested, for example,
    in the fact that the acceleration of a body at
    a given force depends not only on the absolute values of the force
    and velocity, but also on the angle between them.
    In addition, the directions of force and acceleration, in general case, do not coincide.

It is interesting to consider particular cases when the directions of the force and of the acceleration
coincide.
Let us first specify the case when $
\mbox{\boldmath $v$} \perp \mbox{\boldmath $a$} $. Then we have
$$
\mbox{\boldmath $F$} =
 {\frac{m} {\sqrt{1-\beta^{2}}}}
 \mbox{\boldmath $a$} = M\mbox{\boldmath $a$}
 \eqno(16)
$$
and
 $\mbox{\boldmath $F$} \uparrow\uparrow \mbox{\boldmath $a$} $.
  In this case, the measure of inertia is the scalar (one-component)
  quantity--see definition  (5) of the relativistic momentum in
   the relativistic equation of motion (6).
   By obvious reasons, the quantity  $M$ is sometimes called as the transverse mass,
   and sometimes~-- as the Einstein mass, since it appears in the famous Einstein relation
   $E=Mc^{2}$.

In the case when $ \mbox{\boldmath $v$} \uparrow\uparrow
\mbox{\boldmath $a$} $, , formulas (10), (11) lead to the equation
$$
\mbox{\boldmath $F$} =
{\frac{m}{\left(1-\beta^{2}\right)^{3/2}}}\mbox{\boldmath $a$},
\eqno(17)
$$
in this case a one-component quantity ${\frac{m}{\left(1-\beta^{2}\right)^{3/2}}}$
seems to be  a measure of inertia,  sometimes it is called the longitudinal mass.

So, we can distinguish three real applicants for the "status"  \hspace{1mm} of the relativistic mass:
invariant mass $m$, Einstein mass $M$, matrix (tensor) mass $\mu_{ij}$.
We will analyze the advantages and disadvantages of each of the applicants.

The mass $m$ preserves only one  from three features  of the classical mass~--
 recall, these are  invariance, additivity and the measure of inertia~--
in the relativistic dynamics retains only  invariance.
It is convenient for using   in  the micro-world physics, because it helps
 to identify and distinguish   elementary particles. The main "drawback" \hspace{1mm}
 is that the concept of  mass  in relativistic dynamics  is rationalized on the arguments  of relativistic
 invariance only
$$
E^{2}-c^{2}p^{2}=m^{2}c^{4}, \eqno(18)
$$
but not from the form of equation of motion in mechanics.
In other words, the very approach to the definition of the concept of mass is
 dramatically  changed,  and  the  basic  succession  of the concepts of
  the classical and relativistic masses is lost.
  In fact, we have a situation similar to the identification of inertial and gravitational masses,
  although in essence these are different physical entities.
   The  classical mass, in the first place,  is claimed to be a measure
 of inertia, but the relativistic mass~-- as a relativistic invariant related
  to the   four-dimensional momentum (up to the multiplier $1/c$):
$$
m={\frac{1}{c}}{\sqrt{E^{2}-p^{2}c^{2}}} = {\frac{1}{c}}
\sqrt{-p^{2}_{\mu}}. \eqno(19)
$$

Einstein's mass  possesses only one feature -- it is the additivity.
It is a measure of the total energy reserved in the body,  which is conserved.
This mass is involved in  the definition for the  relativistic  momentum,
providing  us with the  basic  succession of the forms for  nonrelativistic and
relativistic equations of motion, expressed in terms of momentum.
Note that it may serve  a measure of inertia for the case of the motion of a relativistic body
along a circle.

The matrix mass  $\mu_{ij}$  inherits from the classical mass the property
of being a measure of inertia. But the main peculiarity is  that it leads to detecting
anisotropic character of the inertial properties of the relativistic body,
which was  hidden when other masses are used.
As a disadvantage, we  can consider a rather complicated character of that  matrix mass
for students' perception,  and that complexity hardly van be overcome by the school children.

So, let us  summarize. Each of the three discussed  senses of the mass
has its advantages and weaknesses.
  Each of the approaches is intrinsically consistent and being applied correctly
  leads to the same physical consequences, verified by experience.
   Therefore, any of them has the equal grounds for existing.
    The choice of the approach depends
     on   the scientific interests of the researcher,
     also on  his personal attitude to the point: which property of the classical
     mass is taken as a  dominant one and which of them  should be preserved in the transition
     to relativistic dynamics.

The most constructive approach, in our opinion, is that one should not focus on any one understanding of
 the relativistic mass. The situation is similar to that in the relativistic kinematics.
 Here they freely  operate with the concepts of the length of a moving and resting body,
 with the time interval between events in an arbitrary reference frame, and with  the interval of
 the proper time. And the do nor argue on the question -- which  length or which  time interval are "real".
 Moreover, the use of different lengths and time intervals turns to be convenient from the point of
 view of correct interpretation of relativistic kinematic effects.
 So, in relativistic dynamics, the simultaneous use of different mass-concepts can contribute
 to the creating  more effective tools in this theory.
 The main thing  is to be able to clearly delineate the  used concepts  and apply them adequately.
 A professional physicist should understand that if we are talking about the mass
 of a newly discovered elementary particle, then of course an invariant mass  $m$ is most appropriate;
 if we discuss  the question of increasing the mass of a spacecraft moving with a near-light
velocity and of the energy costs involved in carrying out such a flight,
then Einstein's mass  $M$ is mostly helpful, and so on.


\begin{thebibliography}{}
\bibitem{Okun}  L.B. Okun. The "relativistic"\hspace{1mm}mug. arXiv: 1010.5400v1 [physics.pop-ph] 26 Oct 2010.
\bibitem{FeynmanL}  Feynman R. P., Leighton R. B. Sands M. The Feynman lectures on physics. Addison-Wesley (1963).
\end{thebibliography}
\end{document}


\begin{center}
\bf{ВТОРОЙ ЗАКОН НЬЮТОНА И РЕЛЯТИВИСТСКАЯ МАССА\\[5mm]}
\end{center}
\begin{center}
В. А. Плетюхов \\
Брестский государственный университет, Брест, Беларусь\\[3mm]
12.02.2018
\end{center}
\hspace{2.0cm}
\begin{abstract}
В настоящей работе обсуждаются различные трактовки понятия массы в релятивистской динамике. Предложен новый способ введения массы, основанный на записи релятивистского уравнения движения в форме второго закона Ньютона. В этом подходе масса выступает не как скаляр, а как тензор. Тензорная масса позволяет просто описать анизотропный характер инертных свойств релятивистского объекта.  \looseness=-1
\end{abstract}
\hspace{1.0cm}

После создания специальной теории относительности прошло уже более 100 лет. По основам теории среди серьёзных специалистов существует полное единство мнений. Тем не менее, по отдельным частным вопросам до сих пор остаются разночтения, которые время от времени порождают жаркие дебаты. В качестве примера можно сослаться на публикацию известного российского физика-теоретика академика Л.Б.~Окуня \cite{Okun}. \looseness=-1

Автор настоящей статьи уже более 40 лет читает курс теории относительности студентам и магистрантам физико-математического факультета Брестского университета. Наша точка зрения на релятивистскую массу отличается от позиции как уважаемого академика, так и критикуемых им оппонентов. Хотелось бы поделиться своими соображениями с читателями, которые интересуется данным вопросом. \looseness=-1

Начнём с самого тривиального. Как известно, классическое уравнение движения для точечного тела имеет вид
$$
{\frac{d}{dt}} (m\mbox{\boldmath $v$}) = \mbox{\boldmath $F$}
\qquad  \text{или} \qquad
{\frac{d \mbox{\boldmath $p$}} {dt}} = \mbox{\boldmath $F$}  \quad (\mbox{\boldmath $p$}=m\mbox{\boldmath $v$}),
\eqno(1)
$$
где $m$ -- инвариантная величина в смысле зависимости от скорости
(а значит и от времени), называемая массой. Предполагается также,
что масса $m$  обладает свойством аддитивности и может
рассматриваться как мера количества вещества, содержащегося в
теле. Независимость   $m$ от времени позволяет переписать (1)
следующим образом:  \looseness=-1
$$
\mbox{\boldmath $F$} = m {\frac{d \mbox{\boldmath $v$}}{dt}} = m\mbox{\boldmath $a$}.
\eqno(2)
$$
Уравнение (2) называется вторым законом Ньютона. Масса $m$  в (2) выступает
 в качестве коэффициента пропорциональности между силой и ускорением.
 В данном качестве она характеризует инертные свойства тела, т.е.
  его способность приобретать определённое ускорение под действием определённой силы.
  Конкретно, масса $m$ численно равна силе, сообщающей телу единичное ускорение.
   И в этом смысле говорят, что масса является мерой инертности тела.
   Часто её так и называют <инертная масса>.

Таким образом, масса в классической механике обладает тремя основными признаками:
 инвариантность, аддитивность, мера инертности,  которые являются её <визитной карточкой> в данной теории.
 Здесь они мирно сосуществуют, дополняя друг друга.

Иначе обстоит дело в релятивистской динамике. Трёхмерное релятивистское \, уравнение движения имеет вид
$$
{\frac{d}  {dt}} \left(
{\frac{m\mbox{\boldmath $v$}}{\sqrt{1-\beta^{2}}}}
\right) =
\mbox{\boldmath $F$}.
\eqno(3)
$$
Или, если ввести в рассмотрение величину
$$
M={\frac {m } {\sqrt{1-\beta^{2}}}}
\eqno(4)
$$
и релятивистский импульс
$$
\mbox{\boldmath $p$} =
{\frac{m\mbox{\boldmath $v$}} {\sqrt{1-\beta^{2}}}} = M\mbox{\boldmath $v$},
\eqno(5)
$$
можно записать (3) в форме, аналогичной (1):
$$
{\frac{d} {dt}}
\left(  M\mbox{\boldmath $v$} \right) = \mbox{\boldmath $F$}, \qquad
{\frac{d \mbox{\boldmath $p$}}{dt}} = \mbox{\boldmath $F$}.
\eqno(6)
$$
Инвариантная величина $m$ в уравнении (3) уже не может рассматриваться как мера инертности. Исходя из этого, некоторая часть физиков (смотри, напр., \cite{FeynmanL}) полагает, что роль массы в релятивистской динамике выполняет величина   (4), замещающая классическую массу $m$  в релятивистском уравнении движения. Однако очевидно, что и величина   $M$ также не является мерой инертности, поскольку в
общем случае $\textbf{F}\neq M\textbf{a}$.

Возникает вопрос, корректно ли вообще говорить о существовании в релятивистской динамике величины, которая может претендовать на роль меры инертности в выше сформулированном смысле. Чтобы ответить на этот вопрос, выясним, можно ли уравнение (3) представить в форме, аналогичной второму закону Ньютона. Для этого возьмём производную, фигурирующую в левой части уравнения (3):
$$
{\frac{d} {dt}}
\left(
{\frac{m \mbox{\boldmath $v$}} {\sqrt{1-\beta^{2}}}}
\right) =
{\frac{m} {\sqrt{1-\beta^{2}}}}
{\frac{d\mbox{\boldmath $v$}} {dt}}
+
{\frac{m\mbox{\boldmath $v$} \left(\mbox{\boldmath $v$}, {\frac{d\mbox{\boldmath $v$}} {dt}} \right) } { c^{2} \left( 1-\beta^{2} \right)^{3/2}}}=
$$
$$
={\frac{m}{\sqrt{1-\beta^{2}}}}
\left(
1+ {\frac
{\mbox{\boldmath $v$}
\cdot
\mbox{\boldmath $v$}
}
{
c^{2} \left( {1-\beta^{2}} \right)
     }
   }
\right)\mbox{\boldmath $a$}  =
{\frac{m}{\sqrt{ 1-\beta^{2} } }}
\left( 1 + {\frac{\mbox{\boldmath $\beta$}\cdot \mbox{\boldmath $\beta$}}   {1-\beta^{2}}}\right)
\mbox{\boldmath $a$}.
\eqno(7)
$$
Здесь $ \mbox{\boldmath $\beta$} = {\frac{\mbox{\boldmath $v$}} {c}}$,
$
\left(\mbox{\boldmath $v$}, {\frac{d\mbox{\boldmath $v$}}  {dt}} \right) =
\left( \mbox{\boldmath $v$},\mbox{\boldmath $a$} \right)$ -- скалярное произведение;
$\mbox{\boldmath $v$} \cdot \mbox{\boldmath $v$}$,
$\mbox{\boldmath $\beta$}\cdot \mbox{\boldmath $\beta$}$ -- прямое (диадное) произведение векторов. Если представить вектора \mbox{\boldmath $F$}  и \mbox{\boldmath $a$}  в виде столбцов
$$
\mbox{\boldmath $F$} =
 \left(\begin{array}{ccc}
F_{1} \\
F_{2} \\
F_{3} \\
\end{array}\right), \qquad
\mbox{\boldmath $a$} =
\left(\begin{array}{ccc}
a_{1} \\
a_{2} \\
a_{3} \\
\end{array}\right),
\eqno(8)
$$
то диада $\mbox{\boldmath $\beta$}\cdot \mbox{\boldmath $\beta$}$  представляет собой матрицу
$$
\mbox{\boldmath $\beta$}\cdot \mbox{\boldmath $\beta$} =
\left(\begin{array}{ccc}
\beta_{1}^{2} &\beta_{1}\beta_{2} &\beta_{1}\beta_{3}  \\
\beta_{2}\beta_{1} & \beta_{2}^{2} &\beta_{2}\beta_{3}    \\
\beta_{3}\beta_{1} & \beta_{3}\beta_{2}    & \beta_{3}^{2}                        \end{array}\right).
\eqno(9)
$$
В преобразованиях (7) использовано свойство диады
 $(\mbox{\boldmath $a$}\cdot \mbox{\boldmath $b$})\mbox{\boldmath $c$} =
 \mbox{\boldmath $a$}(\mbox{\boldmath $b$}, \mbox{\boldmath $c$})$.

С учетом (7) уравнение (3) можно записать в виде
$$
\mbox{\boldmath $F$} = \mu \mbox{\boldmath $a$},
\eqno(10)
$$
где  $\mbox{\boldmath $F$}$ и  $\mbox{\boldmath $a$}$ -- трёхмерные вектор-столбцы вида (8),  $\mu$ -- матрица $3\times3$
$$
\mu = {\frac{m} {\sqrt{1-\beta^{2}}}}
\left( 1 + {\frac{\mbox{\boldmath $\beta$}\cdot \mbox{\boldmath $\beta$}} {1-\beta^{2}}}\right).
\eqno(11)
$$
В индексных обозначениях будем соответственно иметь:
$$
F_{i}=\mu_{ij}a_{j},
\eqno(12)
$$
где
$$
\mu_{ij} =  {\frac{m} {\sqrt{1-\beta^{2}}}}
\left( \delta_{ij} + {\frac{\beta_{i} \beta_{j}}   {1-\beta^{2}}}\right)
\eqno(13)
$$
и по повторяющимся индексам подразумевается суммирование.

Уравнения (10), (12) можно обратить, а именно:
$$
\mbox{\boldmath $a$} = \mu^{-1}\mbox{\boldmath $F$}, \qquad
a_{i}=\mu_{ij}^{-1} F_{j},
\eqno(14)
$$
где обратная матрица $\mu^{-1}$  имеет вид
$$
\mu^{-1} = {\frac{\sqrt{1-\beta^{2}}}{m}}
\left( 1 - {\mbox{\boldmath $\beta$}\cdot \mbox{\boldmath $\beta$}}\right), \qquad
\mu_{ij}^{-1} = {\frac{\sqrt{1-\beta^{2}}}{m}}
\left(
\delta_{ij} - \beta_{i}\beta_{j}
\right).
\eqno(15)
$$

Очевидно, что именно матрица $\mu$ , выступающая в качестве <коэффициента
 пропорциональности> между векторами силы и ускорения, наиболее полно и точно
 отражает смысл понятия <мера инертности> в релятивистской динамике. Матричный характер
 массы как меры инертности означает, что инертные свойства релятивистского тела неизотропны.
 Указанная неизотропность связана с тем, что движущееся тело автоматически создаёт в пространстве
  в каждый момент времени выделенное направление, совпадающее с направлением скорости тела.
   Поэтому можно сказать, что она (неизотропность) носит кинематический характер и проявляется,
    например, в том, что ускорение тела при заданной силе зависит не только от абсолютных
    значений силы и скорости, но и от угла между ними в каждый данный момент времени.
    Кроме того, направления силы и ускорения, вообще говоря, не совпадают.

Интересно рассмотреть частные случаи, когда направления силы и ускорения совпадают.
Возьмём сначала случай, когда
$ \mbox{\boldmath $v$} \perp \mbox{\boldmath $a$} $  . Тогда имеем
$$
\mbox{\boldmath $F$} =
 {\frac{m} {\sqrt{1-\beta^{2}}}}
 \mbox{\boldmath $a$} = M\mbox{\boldmath $a$}
 \eqno(16)
$$
и $\mbox{\boldmath $F$} \uparrow\uparrow \mbox{\boldmath $a$} $.
В этом случае мерой инертности является скалярная (однокомпонентная)
величина  $M$, содержащаяся в определении (5) релятивистского импульса
и релятивистском уравнении движения (6). По понятной причине величину
$M$ иногда называют поперечной массой, иногда же -- эйнштейновской массой, поскольку именно она фигурирует в знаменитом эйнштейновском соотношении $E=Mc^{2}$.

В случае, когда $ \mbox{\boldmath $v$} \uparrow\uparrow \mbox{\boldmath $a$} $,
формулы (10), (11) приводят к уравнению
$$
\mbox{\boldmath $F$} =
{\frac{m}{\left(1-\beta^{2}\right)^{3/2}}}\mbox{\boldmath $a$},
\eqno(17)
$$
то есть в этом случае в качестве меры инертности выступает также
однокомпонентная величина ${\frac{m}{\left(1-\beta^{2}\right)^{3/2}}}$ , называемая иногда продольной массой.

Итак, можно выделить три реальных претендента на <звание> релятивистской массы:
 инвариантная масса $m$, эйнштейновская масса  $M$, матричная (тензорная) масса
  $\mu_{ij}$. Разберём под этим углом зрения преимущества и недостатки каждого из претендентов.  \looseness=-1

Масса $m$ из трёх основных признаков классической массы (напомним,
это~-- инвариантность, аддитивность и мера инертности)
 в релятивистской динамике сохраняет только один~--
 инвариантность. Удобна для использования в физике микромира,
 поскольку позволяет идентифицировать микрочастицу. Основной <недостаток>
 состоит в том, что понятие массы $m$  в релятивистской динамике приходится вводить,
 исходя не из уравнения движения, а из релятивистского соотношения  \looseness=-1
$$
E^{2}-c^{2}p^{2}=m^{2}c^{4},
\eqno(18)
$$
не имеющего прямого отношения к уравнению движения.
Другими словами, принципиально меняется сам подход к определению понятия массы,
 при этом теряется сущностная преемственность понятий классической и релятивистской масс.
 Фактически мы имеем ситуацию, сходную с отождествлением инертной и гравитационной масс,
 хотя по сути -- это разные физические понятия. Аналогично, классическая масса
   позиционируется, в первую очередь, как мера инертности, релятивистская же масса $m$  --
   как модуль четырёхмерного импульса (с точностью до несущественного множителя $1/c$ ):  \looseness=-1
$$
m={\frac{1}{c}}{\sqrt{E^{2}-p^{2}c^{2}}} =
{\frac{1}{c}} \sqrt{-p^{2}_{\mu}}.
\eqno(19)
$$

Эйнштейновской массе $M$  из трёх свойств классической массы также присуще только одно --
 аддитивность. Является мерой запаса полной энергии тела и,
 следовательно, сохраняется. Фигурирует в определении релятивистского импульса,
 обеспечивая преемственность формулировок нерелятивистского и релятивистского
 уравнений движения, выраженных через импульс. Служит мерой инертности при движении
 релятивистского тела по окружности.

Матричная масса $\mu_{ij}$  наследует от классической массы свойство
быть мерой инертности. Но главное -- позволяет обнаружить анизотропный характер
инертных свойств релятивистского тела, что остаётся скрытым при использовании других масс.

Недостатком, впрочем условным, можно считать достаточно сложный характер понятия матричной
массы для восприятия студентами и практически недоступный для школьников.

Итак, подведём итог. Каждая из трёх обсуждаемых масс имеет свои преимущества и слабые
места как с точки зрения преемственности по отношению к понятию классической массы,
так и с точки зрения всестороннего и полного раскрытия физического содержания релятивистской динамики.
 Каждый из подходов внутренне непротиворечив и при правильной, последовательной реализации приводит
 к одинаковым физическим следствиям, проверяемым на опыте. Поэтому любой из них имеет,
 в принципе, право на существование, но без права на абсолютизацию.
 Выбор подхода зависит, в первую очередь, от сферы научных интересов исследователя,
 а также от его личной позиции по поводу того, какое свойство классической массы считать доминантным
 и следует сохранить при переходе к релятивистской динамике.

Самый же конструктивный подход, на наш взгляд, состоит в том, что
не следует зацикливаться на каком-то одном определении релятивистской массы.
В качестве примера можно привести подход, принятый в релятивистской кинематике.
Здесь спокойно оперируют понятиями длины движущегося и покоящегося тела,
 с промежутком времени между событиями в произвольной ИСО и промежутком собственного времени.
  И никто не спорит, какая длина или какой именно промежуток времени являются <настоящими>.
   Более того, использование разных по смыслу длин и промежутков времени оказывается весьма удобным
   с точки зрения правильной трактовки релятивистских кинематических эффектов.
    Так, и в релятивистской динамике одновременное использование различных масс
    может способствовать созданию более эффективного инструментария данной теории.
    Главное при  этом~-- \, уметь чётко разграничивать используемые понятия и адекватно
    их применять. Физику-профессионалу должно быть понятно, что если речь идёт о массе
    вновь открытой элементарной частицы, то конечно подразумевается инвариантная масса;
    если же обсуждается вопрос об увеличении массы космического корабля, движущегося с
    околосветовой скоростью, и об энергетических затратах на осуществление
     такого полёта, то подразумевается эйнштейновская масса  $M$ и т. д.